\documentclass[prl,twocolumn,showpacs,superscriptaddress,floatfix]{revtex4}
 	
 \def\AuthorList{true} 

\usepackage{ifthen}
\usepackage{slashed}
\usepackage{epsfig}
\usepackage{amsmath}
\usepackage{comment}
\usepackage{longtable}
\usepackage{subfigure}
\usepackage{color}

\newcommand{\pythia}{{\sc pythia}}

\newcommand{\GeVM}[1]{#1~GeV/$c^2${}}
\newcommand{\p}[1]{\phantom{#1}}

\newcommand{\MET}{\mbox{$\raisebox{.3ex}{$\not\!$}E_T$}}
\newcommand{\METVEC}{\mbox{$\raisebox{.3ex}{$\not\!$}{\vec E}_T$}}

\begin{document}

\title{Search for the standard model Higgs boson produced in association with top quarks using the full CDF data set}

\ifthenelse{\equal{\AuthorList}{true}}{
  \affiliation{Institute of Physics, Academia Sinica, Taipei, Taiwan 11529, Republic of China}
\affiliation{Argonne National Laboratory, Argonne, Illinois 60439, USA}
\affiliation{University of Athens, 157 71 Athens, Greece}
\affiliation{Institut de Fisica d'Altes Energies, ICREA, Universitat Autonoma de Barcelona, E-08193, Bellaterra (Barcelona), Spain}
\affiliation{Baylor University, Waco, Texas 76798, USA}
\affiliation{Istituto Nazionale di Fisica Nucleare Bologna, $^{ee}$University of Bologna, I-40127 Bologna, Italy}
\affiliation{University of California, Davis, Davis, California 95616, USA}
\affiliation{University of California, Los Angeles, Los Angeles, California 90024, USA}
\affiliation{Instituto de Fisica de Cantabria, CSIC-University of Cantabria, 39005 Santander, Spain}
\affiliation{Carnegie Mellon University, Pittsburgh, Pennsylvania 15213, USA}
\affiliation{Enrico Fermi Institute, University of Chicago, Chicago, Illinois 60637, USA}
\affiliation{Comenius University, 842 48 Bratislava, Slovakia; Institute of Experimental Physics, 040 01 Kosice, Slovakia}
\affiliation{Joint Institute for Nuclear Research, RU-141980 Dubna, Russia}
\affiliation{Duke University, Durham, North Carolina 27708, USA}
\affiliation{Fermi National Accelerator Laboratory, Batavia, Illinois 60510, USA}
\affiliation{University of Florida, Gainesville, Florida 32611, USA}
\affiliation{Laboratori Nazionali di Frascati, Istituto Nazionale di Fisica Nucleare, I-00044 Frascati, Italy}
\affiliation{University of Geneva, CH-1211 Geneva 4, Switzerland}
\affiliation{Glasgow University, Glasgow G12 8QQ, United Kingdom}
\affiliation{Harvard University, Cambridge, Massachusetts 02138, USA}
\affiliation{Division of High Energy Physics, Department of Physics, University of Helsinki and Helsinki Institute of Physics, FIN-00014, Helsinki, Finland}
\affiliation{University of Illinois, Urbana, Illinois 61801, USA}
\affiliation{The Johns Hopkins University, Baltimore, Maryland 21218, USA}
\affiliation{Institut f\"{u}r Experimentelle Kernphysik, Karlsruhe Institute of Technology, D-76131 Karlsruhe, Germany}
\affiliation{Center for High Energy Physics: Kyungpook National University, Daegu 702-701, Korea; Seoul National University, Seoul 151-742, Korea; Sungkyunkwan University, Suwon 440-746, Korea; Korea Institute of Science and Technology Information, Daejeon 305-806, Korea; Chonnam National University, Gwangju 500-757, Korea; Chonbuk National University, Jeonju 561-756, Korea}
\affiliation{Ernest Orlando Lawrence Berkeley National Laboratory, Berkeley, California 94720, USA}
\affiliation{University of Liverpool, Liverpool L69 7ZE, United Kingdom}
\affiliation{University College London, London WC1E 6BT, United Kingdom}
\affiliation{Centro de Investigaciones Energeticas Medioambientales y Tecnologicas, E-28040 Madrid, Spain}
\affiliation{Massachusetts Institute of Technology, Cambridge, Massachusetts 02139, USA}
\affiliation{Institute of Particle Physics: McGill University, Montr\'{e}al, Qu\'{e}bec, Canada H3A~2T8; Simon Fraser University, Burnaby, British Columbia, Canada V5A~1S6; University of Toronto, Toronto, Ontario, Canada M5S~1A7; and TRIUMF, Vancouver, British Columbia, Canada V6T~2A3}
\affiliation{University of Michigan, Ann Arbor, Michigan 48109, USA}
\affiliation{Michigan State University, East Lansing, Michigan 48824, USA}
\affiliation{Institution for Theoretical and Experimental Physics, ITEP, Moscow 117259, Russia}
\affiliation{University of New Mexico, Albuquerque, New Mexico 87131, USA}
\affiliation{The Ohio State University, Columbus, Ohio 43210, USA}
\affiliation{Okayama University, Okayama 700-8530, Japan}
\affiliation{Osaka City University, Osaka 588, Japan}
\affiliation{University of Oxford, Oxford OX1 3RH, United Kingdom}
\affiliation{Istituto Nazionale di Fisica Nucleare, Sezione di Padova-Trento, $^{ff}$University of Padova, I-35131 Padova, Italy}
\affiliation{University of Pennsylvania, Philadelphia, Pennsylvania 19104, USA}
\affiliation{Istituto Nazionale di Fisica Nucleare Pisa, $^{gg}$University of Pisa, $^{hh}$University of Siena and $^{ii}$Scuola Normale Superiore, I-56127 Pisa, Italy}
\affiliation{University of Pittsburgh, Pittsburgh, Pennsylvania 15260, USA}
\affiliation{Purdue University, West Lafayette, Indiana 47907, USA}
\affiliation{University of Rochester, Rochester, New York 14627, USA}
\affiliation{The Rockefeller University, New York, New York 10065, USA}
\affiliation{Istituto Nazionale di Fisica Nucleare, Sezione di Roma 1, $^{jj}$Sapienza Universit\`{a} di Roma, I-00185 Roma, Italy}
\affiliation{Rutgers University, Piscataway, New Jersey 08855, USA}
\affiliation{Texas A\&M University, College Station, Texas 77843, USA}
\affiliation{Istituto Nazionale di Fisica Nucleare Trieste/Udine, I-34100 Trieste, $^{kk}$University of Udine, I-33100 Udine, Italy}
\affiliation{University of Tsukuba, Tsukuba, Ibaraki 305, Japan}
\affiliation{Tufts University, Medford, Massachusetts 02155, USA}
\affiliation{University of Virginia, Charlottesville, Virginia 22906, USA}
\affiliation{Waseda University, Tokyo 169, Japan}
\affiliation{Wayne State University, Detroit, Michigan 48201, USA}
\affiliation{University of Wisconsin, Madison, Wisconsin 53706, USA}
\affiliation{Yale University, New Haven, Connecticut 06520, USA}

\author{T.~Aaltonen}
\affiliation{Division of High Energy Physics, Department of Physics, University of Helsinki and Helsinki Institute of Physics, FIN-00014, Helsinki, Finland}
\author{B.~\'{A}lvarez~Gonz\'{a}lez$^z$}
\affiliation{Instituto de Fisica de Cantabria, CSIC-University of Cantabria, 39005 Santander, Spain}
\author{S.~Amerio}
\affiliation{Istituto Nazionale di Fisica Nucleare, Sezione di Padova-Trento, $^{ff}$University of Padova, I-35131 Padova, Italy}
\author{D.~Amidei}
\affiliation{University of Michigan, Ann Arbor, Michigan 48109, USA}
\author{A.~Anastassov$^x$}
\affiliation{Fermi National Accelerator Laboratory, Batavia, Illinois 60510, USA}
\author{A.~Annovi}
\affiliation{Laboratori Nazionali di Frascati, Istituto Nazionale di Fisica Nucleare, I-00044 Frascati, Italy}
\author{J.~Antos}
\affiliation{Comenius University, 842 48 Bratislava, Slovakia; Institute of Experimental Physics, 040 01 Kosice, Slovakia}
\author{G.~Apollinari}
\affiliation{Fermi National Accelerator Laboratory, Batavia, Illinois 60510, USA}
\author{J.A.~Appel}
\affiliation{Fermi National Accelerator Laboratory, Batavia, Illinois 60510, USA}
\author{T.~Arisawa}
\affiliation{Waseda University, Tokyo 169, Japan}
% ...

\collaboration{CDF Collaboration\footnote[2]{With visitors from
$^a$Istituto Nazionale di Fisica Nucleare, Sezione di Cagliari, 09042 Monserrato (Cagliari), Italy,
$^b$University of California Irvine, Irvine, CA 92697, USA,
$^c$University of California Santa Barbara, Santa Barbara, CA 93106, USA,
$^d$University of California Santa Cruz, Santa Cruz, CA 95064, USA,
$^e$Institute of Physics, Academy of Sciences of the Czech Republic, Czech Republic,
$^f$CERN, CH-1211 Geneva, Switzerland,
$^g$Cornell University, Ithaca, NY 14853, USA,
$^h$University of Cyprus, Nicosia CY-1678, Cyprus,
$^i$Office of Science, U.S. Department of Energy, Washington, DC 20585, USA,
$^j$University College Dublin, Dublin 4, Ireland,
$^k$ETH, 8092 Zurich, Switzerland,
$^l$University of Fukui, Fukui City, Fukui Prefecture, Japan 910-0017,
$^m$Universidad Iberoamericana, Mexico D.F., Mexico,
$^n$University of Iowa, Iowa City, IA 52242, USA,
$^o$Kinki University, Higashi-Osaka City, Japan 577-8502,
$^p$Kansas State University, Manhattan, KS 66506, USA,
$^q$Ewha Womans University, Seoul, 120-750, Korea,
$^r$University of Manchester, Manchester M13 9PL, United Kingdom,
$^s$Queen Mary, University of London, London, E1 4NS, United Kingdom,
$^t$University of Melbourne, Victoria 3010, Australia,
$^u$Muons, Inc., Batavia, IL 60510, USA,
$^v$Nagasaki Institute of Applied Science, Nagasaki, Japan,
$^w$National Research Nuclear University, Moscow, Russia,
$^x$Northwestern University, Evanston, IL 60208, USA,
$^y$University of Notre Dame, Notre Dame, IN 46556, USA,
$^z$Universidad de Oviedo, E-33007 Oviedo, Spain,
$^{aa}$CNRS-IN2P3, Paris, F-75205 France,
$^{bb}$Texas Tech University, Lubbock, TX 79609, USA,
$^{cc}$Universidad Tecnica Federico Santa Maria, 110v Valparaiso, Chile,
$^{dd}$Yarmouk University, Irbid 211-63, Jordan.
}}
\noaffiliation
 
}

\ifthenelse{\equal{\AuthorList}{false}}{
\begin{figure}  
  \leftline{\includegraphics[scale=0.5]{figures/cdfii_thumb_logo.eps}\hfill
    CDF/PUB/EXOTIC/CDFR/10899}
\end{figure}

\vspace*{0.5cm} 

\author{The CDF Collaboration}
\affiliation{URL http://www-cdf.fnal.gov}
}

\date{\today}

\begin{abstract}
A search is presented for the standard model Higgs boson produced in association with top quarks using the full Run II proton-antiproton collision data set, corresponding to 9.45 fb$^{-1}$, collected by the Collider Detector at Fermilab.  No significant excess over the expected background is observed, and 95\% credibility-level upper bounds are placed on the cross section $\sigma (t\bar{t} H \to \text{lepton} + \text{missing transverse energy} + \text{jets})$.  For a Higgs boson mass of \GeVM{125}, we expect to set a limit of 12.6, and observe a limit of 20.5 times the standard model rate.  This represents the most sensitive search for a standard model Higgs boson in this channel to date.
\end{abstract}



\maketitle

The mechanism of electroweak symmetry breaking~\cite{higgs} in the standard model (SM)~\cite{sm} predicts the existence of a massive particle called the Higgs boson.  The CDF and D0 collaborations have reported evidence for a particle consistent the SM Higgs boson with a mass between 120 and 135~GeV$/c^2$ produced in association with a $W$ or $Z$ boson with decays to two $b$ quarks~\cite{Aaltonen:2012qt}.   The CMS and ATLAS collaborations have reported the observation of a particle consistent with the SM Higgs boson with a mass of approximately \GeVM{125}, which decays to two photons, two $W$ bosons, or two $Z$ bosons~\cite{higgs_discovery}.  Many other predicted couplings of the SM Higgs boson are currently neither observed nor excluded.  In the SM, the fermion masses are generated by Yukawa couplings between the Higgs and the fermion fields with coupling strength proportional to the fermion masses.  As the most massive known fermion, the top quark is expected to couple most strongly to the Higgs boson, which consequently may be produced relatively more abundantly in association with a top quark pair, via radiation or top-quark fusion~\cite{tthxsec1,tthxsec2}.  Samples of top-quark pair events with a few percent-level contamination from other processes can be selected at CDF~\cite{topmass}, offering smaller background uncertainties than in searches for the SM Higgs boson produced in association with a vector boson~\cite{cdfbb}.  Hence, the top-quark pair associated production channel provides an important contribution to SM Higgs boson physics.  Furthermore, proposed extensions to the SM could significantly enhance the coupling between the top quark and the Higgs boson~\cite{bsmtth}. This enhancement might allow the observation of a non-SM Higgs boson in this search before reaching sensitivity to a SM Higgs boson, and could help to distinguish a candidate Higgs boson in other searches from the SM Higgs boson.

This Letter reports a search for the SM Higgs boson produced in association with top quarks ($t\bar{t}H$).  We utilize the full data set recorded with the CDF II detector.  The data set consists of proton-antiproton collisions at a center-of-mass energy of $\sqrt{s} = 1.96$~TeV, and corresponds to an integrated luminosity of 9.45~fb$^{-1}$.  The analysis described in this Letter extends and enhances a previous CDF search which used 319 pb$^{-1}$~\cite{oldtth}, through a vastly increased data set, greater signal acceptance and improved background discrimination.

The CDF II detector is a general-purpose particle detector described in Ref.~\cite{Acosta:2004yw}.  It consists of a combined silicon and drift chamber tracking system with a large volume immersed in the 1.4~T field of a solenoid magnet~\cite{silicon, cot}, lead- and iron-scintillator sampling calorimeters~\cite{em, had}, and charged particle detectors outside the calorimeter, which are used to identify muons~\cite{muons}.  A right-handed cylindrical coordinate system is used with the origin in the center of the detector, with $\theta$ and $\phi$ denoting the polar and azimuthal angles, respectively.  Pseudorapidity is defined as $\eta \equiv - \ln \tan (\theta / 2)$, and transverse energy and momentum are $E_T \equiv E \sin \theta$ and $p_T \equiv p \sin \theta$, where $E$ and $p$ are the energy and momentum, respectively.

The decay of a pair of top quarks is expected to generate almost exclusively two $W$ bosons and two $b$ quarks.  The $W$ bosons may then decay to lepton-neutrino pairs, or pairs of quarks.  We select events consistent with one leptonic and one hadronic $W$ boson decay by requiring the presence of a single reconstructed lepton (electron or muon), missing transverse energy (\MET)~\cite{MET}, and four or more calorimeter energy clusters (jets).  At least two of the jets in each event are required to be consistent with the fragmentation of a $b$ quark ($b$-tagged).  Because a low-mass ($m_H \leq$~\GeVM{135}) SM Higgs boson is expected to decay mostly to pairs of $b$ quarks, or pairs of $W$ bosons, that will decay predominantly to pairs of $u, d, s$, or $c$ quarks, large $b$-tag and jet multiplicities are requested by the selection. Approximately 90\% of the selected search sample is composed of top-quark pairs, with the remainder consisting of $W$ or $Z$ bosons accompanied by jets ($W/Z + \text{jets}$), single top-quarks, dibosons, and strong force mediated (QCD)  multijets.  Table~\ref{acceptance} shows the expected composition of the data sample.

To select events during data taking we require the presence of a charged lepton (electron $e$ or muon $\mu$) candidate with transverse momentum $p_T \ge 18$~GeV/$c$.  We further require that the lepton candidate satisfies identification quality requirements as in Ref.~\cite{topmass}.  We require that $\MET{}$ be greater than 10~GeV, 20~GeV, or 25~GeV in events containing a muon candidate, an electron candidate satisfying $|\eta| \le 1.1$, and an electron candidate satisfying $|\eta| > 1.1$, respectively.  These \MET{} requirements are chosen to optimize the signal selection efficiency and the rejection of instrumental backgrounds, which differ in the three samples.  Jets are reconstructed using a cone-based clustering algorithm, with a cone radius ($R = \sqrt{\Delta \phi^2 + \Delta \eta^2}$) of 0.4~\cite{jetclu}.  Jet energies are corrected for instrumental effects~\cite{Bhatti:2005ai}, and the corrected jets are required to have $E_T > 20$~GeV and $\left|\eta\right| < 2.0$.  We use two different algorithms to tag $b$ jets as in Ref.~\cite{Aaltonen:2011rt}.  One algorithm relies on the reconstruction of secondary decay vertices from long-lived hadrons within the jet cone~\cite{secvtx}, while the other estimates the likelihood that not all tracks in the jet cone intersect the beam line~\cite{jetprob}.  Jets identified by either algorithm are considered as tagged, offering higher tagging efficiency than obtained by the use of one algorithm alone.

We model the various backgrounds using a combination of Monte Carlo (MC) simulation and data.  We simulate the $t \bar{t}$, diboson, $W/Z + \text{jets}$, and single-top backgrounds using the \textsc{powheg}~\cite{powheg}, \textsc{pythia}~\cite{pythia}, \textsc{alpgen}~\cite{Mangano:2002ea} and \textsc{MadEvent}~\cite{madgraph} generators, respectively.  We model the QCD multijet background using a data-driven model~\cite{topmass}.  For backgrounds involving top quarks, we have used $m_t = $~\GeVM{172.5}.  Signal models are generated by \textsc{pythia}, with Higgs boson masses in \GeVM{5} increments in the range 100 $\le m_H \le $~\GeVM{150}.  The CTEQ5L parton distribution functions~\cite{CTEQ5L} and a detailed simulation of the response of the CDF II detector using {\sc geant}3~\cite{GEANT3} is employed in all Monte Carlo samples.

The search sample is subdivided into independent categories of different expected signal-to-background ratio and background composition to maximize the search sensitivity~\cite{jswthesis}. Under the selection requirements described above, the reconstructed jet multiplicity spectrum in $t \bar{t} H$ events peaks at five jets, while the reconstructed jet multiplicity spectrum for $t \bar{t}$ peaks at four jets.  Hence, we separate events with four, five, or six or more jets.  The jet multiplicity samples are then separated by $b$-tag multiplicity.  The events with six or more jets, at least three of which are $b$-tagged, feature the largest expected signal-to-background ratio and provide the most sensitivity for a low-mass Higgs boson.  

\begin{table*}
 \caption{Expected number of events from the various processes composing our data sample, requiring two or more $b$ tags, with background rates and uncertainties taken from the posterior likelihoods.  Uncertainties shown are correlated.  Signal yields are quoted assuming $m_H$ = 125~GeV/$c^2$.\label{acceptance}}
\begin{center}
 \begin{ruledtabular}
 \begin{tabular}{lccc}
  Process            &  4 jets                      & 5 jets                      & $\ge 6$ jets                \\ \hline \noalign{\smallskip}
          $t\bar{t}+\text{jets}$ & $\p{..0}962\p{.00} \pm \p{0}89\p{.000}$ & $\p{..}294\p{.000} \pm \p{}27\p{.000}$ & $\p{..0}77\p{.00} \pm \p{0}7.1\p{00}$ \\
          $t\bar{t}+b\bar{b}$    & $\p{..00}32\p{.00} \pm \p{0}27\p{.000}$ & $\p{..0}17\p{.000} \pm \p{}14\p{.000}$ & $\p{..00}8.2\p{0} \pm \p{0}6.9\p{00}$ \\
          $W/Z+\text{jets}$      & $\p{..0}105\p{.00} \pm \p{0}32\p{.000}$ & $\p{..0}26\p{.000} \pm \p{0}8.0\p{00}$ & $\p{..00}7.1\p{0} \pm \p{0}2.2\p{00}$ \\
            Multijet             & $\p{..00}31\p{.00} \pm \p{0}16\p{.000}$ & $\p{..00}0.0\p{00} \pm \p{0}1.0\p{00}$ & $\p{..00}0.0\p{0} \pm \p{0}1.0\p{00}$ \\
          Single top             & $\p{..00}19\p{.00} \pm \p{00}2.2\p{00}$ & $\p{..00}3.7\p{00} \pm \p{0}0.43\p{0}$ & $\p{..00}0.61\p{} \pm \p{0}0.070\p{}$ \\
             Diboson             & $\p{..000}5.2\p{0} \pm \p{00}0.44\p{0}$ & $\p{..00}1.2\p{00} \pm \p{0}0.11\p{0}$ & $\p{..00}0.25\p{} \pm \p{0}0.025\p{}$ \\ \hline \noalign{\smallskip}
    Total background             & $\p{..}1150\p{.00} \pm \p{}106\p{.000}$ & $\p{..}340\p{.000} \pm \p{}33\p{.000}$ & $\p{..}93\p{.00} \pm \p{}11\p{.00}$ \\ \hline \noalign{\smallskip}
            Observed             & $  \p{}1133\p{.00}\p{ \pm 000}\p{.000}$ & $  \p{}368\p{.000}\p{ \pm 00}\p{.000}$ & $ \p{}114\p{.00}\p{ \pm 00}\p{.000}$ \\ \hline
         $t\bar{t}H$             & $\p{..000}0.65\p{} \pm \p{00}0.075\p{}$ & $\p{..00}1.1\p{00} \pm \p{0}0.13\p{0}$ & $\p{..00}1.2\p{0} \pm \p{0}0.14\p{0}$ \\
                $WH$             & $\p{..000}0.52\p{} \pm \p{00}0.061\p{}$ & $\p{..00}0.07\p{0} \pm \p{0}0.008\p{}$ & negligible \\
                $ZH$             & $\p{..000}0.09\p{} \pm \p{00}0.011\p{}$ & $\p{..00}0.02\p{0} \pm \p{0}0.002\p{}$ & negligible \\  
 \end{tabular}
 \end{ruledtabular}
\end{center}
\end{table*}

\begin{figure}
\includegraphics[width=0.9\linewidth]{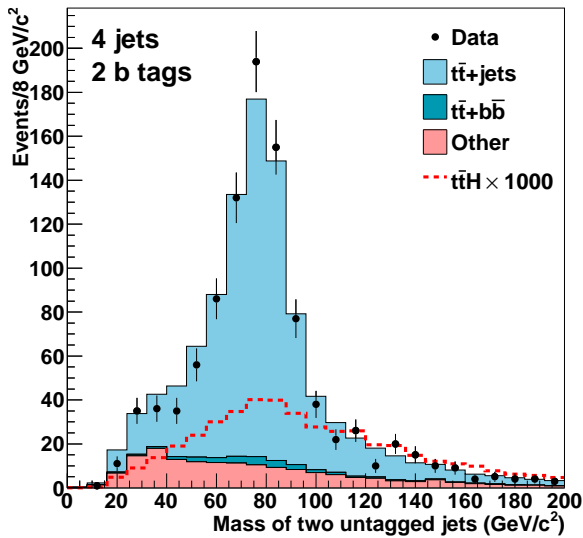}
\caption{Invariant mass of the two jets without $b$ tags, in events containing exactly four jets and exactly two $b$ tags.  The peak of the distribution is consistent with hadronic decays of the $W$ boson.  The effect of systematic uncertainties is not shown.  In the signal model shown, a Higgs boson of $m_{H} = \mbox{\GeVM{125}}$ is assumed.\label{inputvarsMW}}
\end{figure}

\begin{figure}
\includegraphics[width=0.9\linewidth]{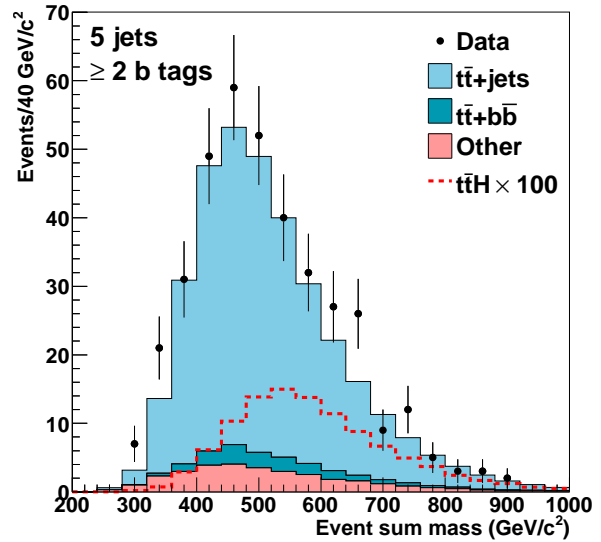}
\caption{The mass of the vector sum of the four-momenta of the identified charged lepton, the neutrino, and all reconstructed jets in events with exactly five jets and at least two $b$ tags.  The effect of systematic uncertainties is not shown.  In the signal model shown, a Higgs boson of $m_{H} = \mbox{\GeVM{125}}$ is assumed.\label{inputvarsESM}}
\end{figure}

\begin{figure}
\includegraphics[width=0.9\linewidth]{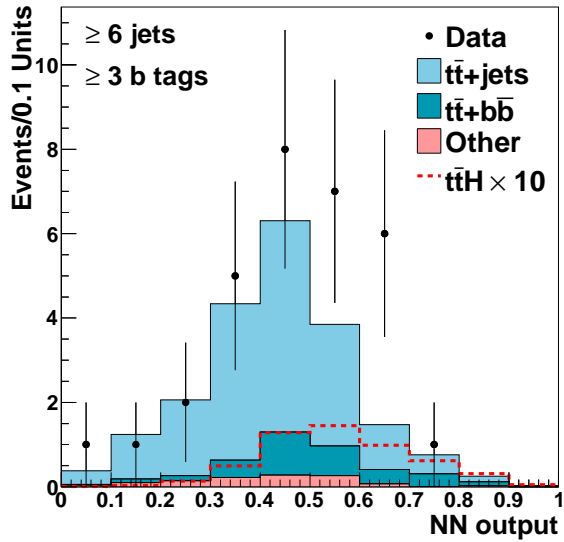}
\caption{The output distribution for the discriminant optimized for the $m_H = \mbox{\GeVM{125}}$ hypothesis, for events with six or more jets and three or more $b$ tags.  The effect of systematic uncertainties is not shown.  In the signal model shown, a Higgs boson of $m_{H} = \mbox{\GeVM{125}}$ is assumed.\label{discr}}
\end{figure}

After defining our search sample, we enhance the isolation of a SM Higgs signal using artificial neural networks (NN)~\cite{Peterson:1993nk}.  Each neural network is trained to separate simulated Higgs signal events from background, with individual networks optimized for each Higgs boson mass hypothesis in each of the previously-described event categories. Each network uses 18 input variables used to discriminate the Higgs boson signal from the backgrounds. These variables are: missing transverse energy, maximum jet $E_T$, second largest jet $E_T$, third largest jet $E_T$, maximum $E_T$ among $b$-tagged jets, mean jet $E_T$, invariant mass of the combination of all objects (jets, lepton, \MET), vector sum of the transverse energies of all objects, scalar sum of the transverse energies of all objects, scalar sum of the transverse energies of all jets, number of energy clusters with $E_T$ between 12 and 20~GeV, minimum separation in $\eta$--$\phi$ space between $b$-tagged jets, separation in azimuth between the lepton and the missing transverse energy, transverse mass of the lepton and missing transverse energy~\cite{Wmt}, mass of the vector sum of the lepton and nearest jet in $\eta$--$\phi$ space, minimum mass of the vector sum of any pair of jets, mass of the vector sum of the two non $b$-tagged jets with the largest $E_T$, and mass of the vector sum of the two $b$-tagged jets with the largest $E_T$.  The modeling of the input distributions has been validated in the subset of the data with only four jets and only two $b$ tags, which is expected to contain a negligible number of signal events relative to the background yield.  Two of these distributions can be seen in Figs.~\ref{inputvarsMW} and~\ref{inputvarsESM}, and the output of the discriminant trained to identify a Higgs boson of mass \GeVM{125} is shown in Fig.~\ref{discr}.

We consider several sources of systematic uncertainty that affect the rate of the involved processes and the shape of the discriminant distributions.  Due to the high jet and $b$-tag multiplicities considered, the dominant systematic uncertainties are associated with estimates of the $b$-tag efficiency and the jet energy scale.  These affect both the rates and the discriminant shapes, and we estimate the effects by independently varying the estimated $b$-tag efficiency and the jet-energy scale within one standard deviation.  These variations in jet-energy scale and tagging efficiency alter the expected acceptance for signal and background by between 1 and 20\%, depending on the selection category.   In addition, to account for uncertainties on the theoretical cross sections of background processes, we assume the following systematic uncertainties on the normalization of simulated backgrounds: 6\% for diboson production, 6\% for single top quark production, 10\% for $t\bar{t}H$ production, and 40\% for $W/Z+\text{jets}$~\cite{Wjetsyst,PhysRevLett.104.131801,Cacciari:2003fi,StopSyst}    Smaller uncertainties include those on the amount of initial- and final-state radiation, parton-distribution function choice, the probability to $b$-tag light-quark jets, and a 6\% uncertainty on the measurement of the integrated luminosity~\cite{jswthesis, lumi}.

No measurement is available of the cross section for top-quark production with additional $b$ quarks generated from QCD radiation.  The next-to-leading-order corrections to leading-order calculations of the production rate of top-quark pairs with additional $b$ quarks have been estimated to be on the order of a factor of two in some regions of phase space~\cite{Bredenstein:2010je}.  To account for this unknown and potentially large systematic uncertainty, inclusive $t\bar{t}$ simulated events were separated into subsamples with additional $b$ quarks generated from QCD radiation ($t\bar{t}+b\bar{b}$), and without ($t\bar{t}+\text{jets}$).  We assume an uncertainty of 10\% on the normalization of the $t\bar{t}+\text{jets}$ component and assume an uncertainty of 100\% on the normalization of the $t\bar{t}+b\bar{b}$ component.
We estimate the effect of individual systematic uncertainties by calculating the expected exclusion sensitivity considering all uncertainties, and then comparing this value to that derived by considering all but one uncertainty. The uncertainty due to the jet-energy scale, $b$-tag efficiency, inclusive top pair cross section, and potential next-to-leading-order effects for $t\bar{t}+b\bar{b}$ individually degrade the expected exclusion sensitivity of the analysis by 7.8\%, 5.4\%, 6.9\%, and 9.0\%, respectively.

\begin{figure}
\includegraphics[width=0.9\linewidth]{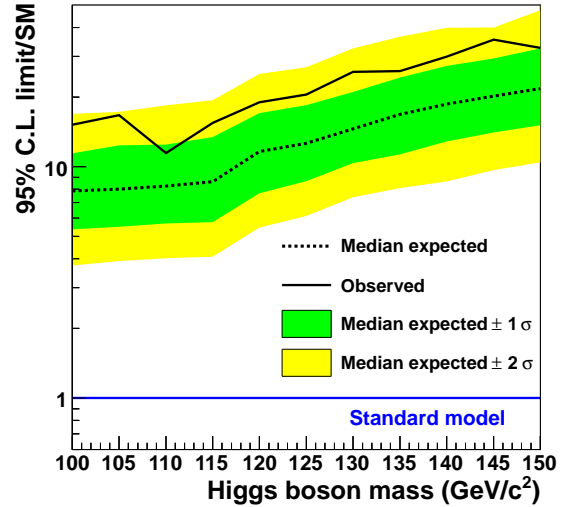}
\caption{Expected and observed 95\% C.L. upper limit as a function of Higgs boson mass for 100 $\le m_H \le$ \GeVM{150}.\label{limits}}
\end{figure}

We compare the distribution of discriminant output observed in data to that of the expected background model.  Observing no evidence for Higgs boson production in the discriminant distributions, we calculate a Bayesian 95\% credibility level (C.L.) limit for each mass hypothesis using the combined binned likelihood of the NN output distributions.  Each of the three jet-multiplicity categories are subdivided into five independent tagging categories. A posterior density is obtained by  multiplying this likelihood by Gaussian prior densities for the background normalizations and systematic uncertainties, leaving the cross section $\sigma (t\bar{t}H \to \ell + \mbox{\MET} + \text{jets})$ with a uniform prior density, with priors truncated to prevent negative predictions.  A 95\% C.L. limit is determined such that 95\% of the posterior density for the cross section accumulates below the limit~\cite{pdgstats}.  The expected limits with one and two standard deviation uncertainty bands and the observed limits are shown as a function of assumed Higgs boson mass in Fig.~\ref{limits}.  Because none of the discriminant function input variables acts as an estimator for the reconstructed Higgs boson mass, the upper credibility limits at different candidate Higgs boson masses are strongly correlated. An excess in the data produces an observed limit that exceeds the expected limit at all masses, at a level of approximately one standard deviation compared to the background-only hypotheses.

In conclusion, we have presented a search for a SM Higgs boson produced in association with a pair of top quarks, in a final state involving a lepton, missing transverse energy, jets, and $b$-tagged jets.  For a Higgs boson mass of \GeVM{125}, we expect a limit of 12.6 and observe a limit of 20.5 times the SM rate, which represents agreement with the background-only prediction at the level of approximately one standard deviation.  The introduction of neural networks and other improvements to the techniques employed in this analysis produce a factor of 17 improvement in sensitivity over the previous search in this channel at CDF~\cite{oldtth} and make this analysis the most sensitive search for $t \bar{t} H$ to date.

We thank the Fermilab staff and the technical staffs of the
participating institutions for their vital contributions. This work
was supported by the U.S. Department of Energy and National Science
Foundation; the Italian Istituto Nazionale di Fisica Nucleare; the
Ministry of Education, Culture, Sports, Science and Technology of
Japan; the Natural Sciences and Engineering Research Council of
Canada; the National Science Council of the Republic of China; the
Swiss National Science Foundation; the A.P. Sloan Foundation; the
Bundesministerium f\"ur Bildung und Forschung, Germany; the Korean
World Class University Program, the National Research Foundation of
Korea; the Science and Technology Facilities Council and the Royal
Society, UK; the Russian Foundation for Basic Research; the Ministerio
de Ciencia e Innovaci\'{o}n, and Programa Consolider-Ingenio 2010,
Spain; the Slovak R\&D Agency; the Academy of Finland; and the
Australian Research Council (ARC).




\begin{thebibliography}{38}
\bibitem[{MET()}]{MET}
\bibinfo{note}{The calorimeter missing $E_T$ ($\METVEC(\mbox{cal})$) is defined
  by the sum over calorimeter towers, $\METVEC(\mbox{cal}) = - \sum_{i} E_T^i
  \hat{n}_i$, where $i$ is calorimeter tower number with $|\eta| < 3.6$,
  $\hat{n}_i$ is a unit vector perpendicular to the beam axis and pointing at
  the $i$th calorimeter tower. The reconstructed missing transverse energy,
  $\METVEC$, is derived by subtracting from $\METVEC(\mbox{cal})$ components of
  the event not registered by the calorimeter, such as muons and jet energy
  adjustments. $\MET$ is the scalar magnitude of $\METVEC$.}
\end{thebibliography}
\end{document}